\documentclass[conference]{IEEEtran}
\usepackage[latin9]{inputenc}
\usepackage{units}
\usepackage{amstext}
\usepackage{graphicx}

\makeatletter
\ifCLASSINFOpdf
\else
\fi
\usepackage{cite}

\makeatother

\begin{document}

\title{Coherent transceiver architecture enabling data transmission and optical
identification }
\author{\IEEEauthorblockN{S. Civelli} \IEEEauthorblockA{CNR-IEIIT\\
Pisa, Italy\\
stella.civelli@cnr.it} \and \IEEEauthorblockN{M. Secondini} \IEEEauthorblockA{Sant'Anna University\\
Pisa, Italy\\
m.secondini@santannapisa.it}\and \IEEEauthorblockN{P. Nadimi Goki} \IEEEauthorblockA{Sant'Anna University\\
Pisa, Italy\\
p.nadimigoki@santannapisa.it}\and \IEEEauthorblockN{L. Pot�} \IEEEauthorblockA{CNIT and Universitas Mercatorum\\
Italy\\
luca.poti@cnit.it} }

\maketitle

%




\begin{abstract}
We propose a coherent transceiver architecture able to transmit information
and enhance the security of the optical network by identifying other
optical systems and sub-systems. Simulations show that identification
is obtained with sufficient reliability---probability of false positive
and negative below $10^{-10}$---in standard operating conditions.
\end{abstract}



\IEEEpeerreviewmaketitle{}

\section{Introduction}

Recently, physical layer security (PLS) has gained a lot of attention
to enhance the overall security of the optical network. Different
PLS techniques are currently investigated including quantum key distribution,
steganography, physical unclonable function (PUF)-based approaches,
and optical identification (OI). The latter consists in assigning
a signature to each system or sub-system of the optical network, devising
a physical characteristic which can be read and translated into a
digital signature with a procedure that ensures uniqueness and unclonability
with good reliability \cite{du2017unclonable,poti2022fitce,goki2023optical}.
The signature of a sub-system can be used for different security purposes
including identification, authentication and monitoring. Recently,
we applied OI taking advantage of the imperfections of the optical
fiber which cause Rayleigh backscattering. The fiber is a PUF with
the Rayleigh backscattering pattern (RBP) obtained when stimulated
by light being its response \cite{du2017unclonable}. The coherent
optical frequency domain reflectometry (C-OFDR) allows to read the
RBP while maintaining its unclonability \cite{COFDR,du2017unclonable,goki2023optical}.
OI using C-OFDR and RBP allows an user---Alice, in the following---to
identify an optical sub-system---Bob, in the following---simply
through its fiber pigtail, without the need of providing it with an
additional device, as for other PUF-based approaches \cite{shamsoshoara2020survey,Mahdian2024photonic}.
In this work, we propose a novel coherent transceiver (C-TRX) architecture,
very similar to a conventional one, which allows to suspend for a
short time data transmission to perform remote user OI with good reliability
thus improving the link security.

\section{Optical identification using C-OFDR}

\begin{figure}
\includegraphics[width=0.5\textwidth]{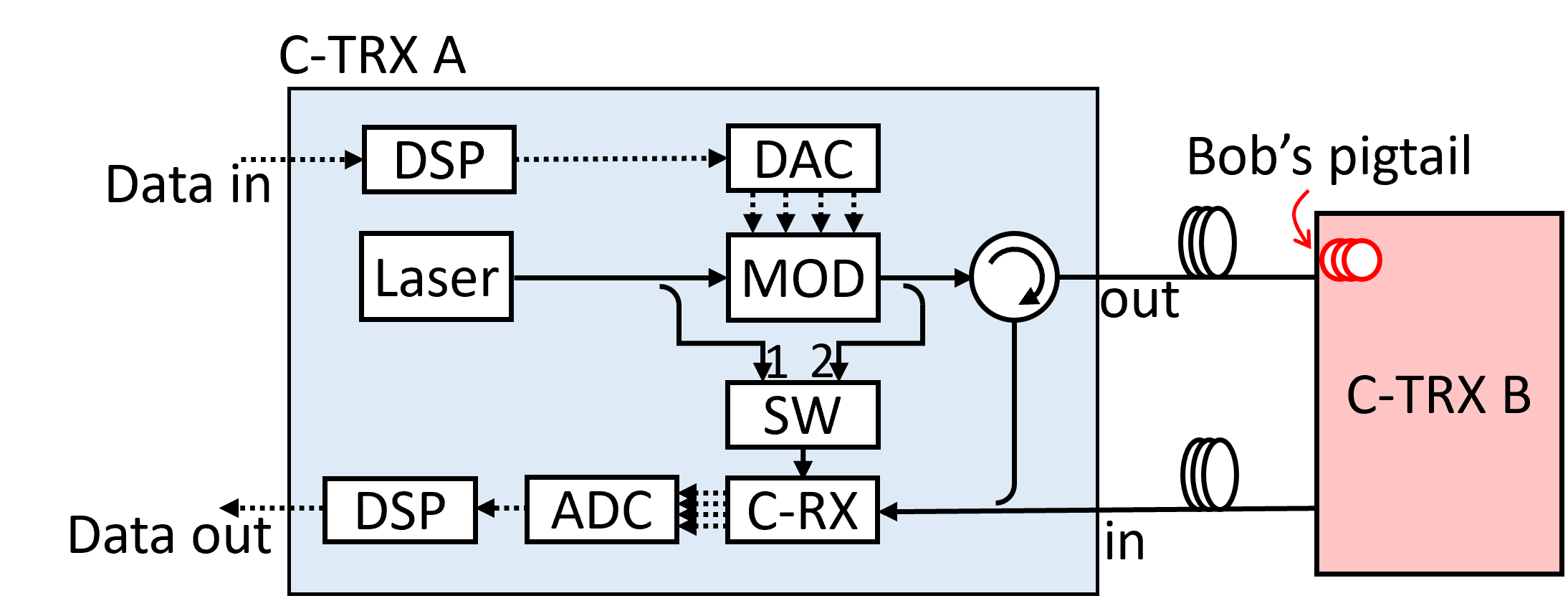}

\caption{\label{fig:TRX}Alice (C-TRX A) can either transmit information or
identify Bob (C-TRX B) through its pigtail using C-OFDR.  DAC: digital-to-analog
converter.}
\end{figure}
OI using C-OFDR is as follows. Alice interrogates Bob's fiber pigtail
of length $L$ with a linear frequency sweep that covers a frequency
range of $\Delta F$ in the sweep time $T_{sw}$ and with rate $\gamma=\Delta F/T_{sw}$.
The backreflected light is coupled with the local oscillator (LO),
and the RBP is acquired with a balanced photodetector (BPD) and a
2-level ADC. The RBP characterizes the fiber, and, in particular,
its spectrum has peaks at the frequencies $f_{k}=\gamma2(d+p_{k})/v$,
with $d$ being the distance between Alice and Bob's pigtail, $0\leq p_{k}\leq L$
being the position of a scattering point in Bob's pigtail, and $v$
being the speed of light in the fiber. Alice digitally filters the
received RBP with bandwidth $B=2L\gamma/v\leq\Delta F$ and  $N=4\Delta FL/v$
samples to extract the RBP corresponding to Bob's fiber pigtail, which
is its signature ($N$-long vector of bits). The signature is compared
with Bob's \emph{ideal} and known signature in terms of Hammin distance
to assess the true identity of the interrogated user \cite{goki2023optical,civelli2023ecocqkd}.
Considering the impact of shot and electronic noise, and assuming
a $\unit[3]{dB}$ splitter for the LO, the overall signal-to-noise
ratio (SNR) on the RBP can be estimated as
\begin{equation}
\mathrm{SNR}=\frac{RP^{2}e^{-2\alpha d}R_{\text{RB}}/2}{qPB+R\mathrm{NEP}^{2}B},\label{eq:SNR}
\end{equation}
where $R$ and $\text{NEP}$ are the responsivity and the noise equivalent
power of the BPD, $P$ the launch power, $\alpha$ the fiber attenuation,
$R_{\text{RB}}$ the overall reflectivity of the pigtail, and $q$
the electron charge. Note that, since the SNR depends linearly on
 $T_{sw}$, repeating the measure $N_{sw}$ times is equivalent to
consider a longer sweep time of $N_{sw}T_{sw}$. Since the spatial
resolution of C-OFDR is $v/2/\Delta F$, to improve the resolution
of C-OFDR for sensing and monitoring purposes, a tunable laser  is
commonly used to increase the frequency range $\Delta F$ \cite{du2017unclonable}.
 However, as we will show in following, a smaller frequency range,
achievable with a conventional C-TRX, is sufficient for OI.

\section{Transceiver architecture}

In this paper, we propose a C-TRX architecture, a conventional C-TRX
with minor modifications, for both data transmission and OI, when
the link is bidirectional. The proposed C-TRX, described in Fig.\,\ref{fig:TRX},
is a standard C-TRX with one additional circulator, one coupler, and
one switch (SW). In transmission mode, the proposed C-TRX---C-TRX
A, Alice---sends information to C-TRX B through its output port,
and receive information through its input port. In transmission mode,
the modulator (MOD) encodes information data in a standard manner.
The first port of SW is close, while the second   is open, allowing
the LO (before MOD) to reach the coherent receiver (C-RX) and perform
coherent detection of the signal coming from the input port.  In 
OI mode, the architecture allows Alice to identify C-TRX B through
its pigtail, as follows. Alice sends a frequency sweep, generated
by laser and MOD, in the output fiber to reach C-TRX B. The first
port of SW is open, while the second is close, such that the LO after
MOD, with the frequency sweep, reaches  C-RX. The backreflected light
from Bob's pigtail passes through the circulator and arrives to the
C-RX, which detects the RBP. 

\section{System setup, specifications and reliability}

\begin{figure}
\centering\includegraphics[width=0.5\textwidth]{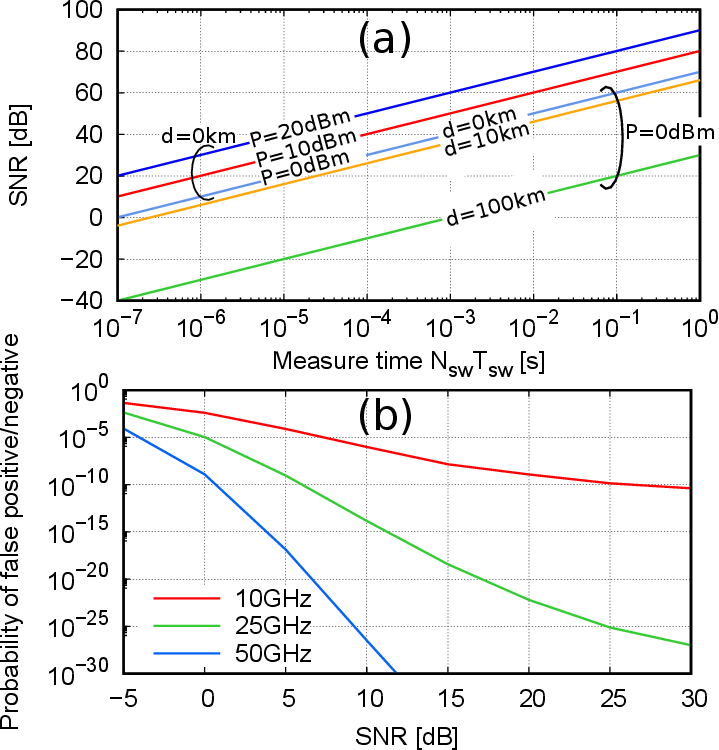}

\caption{\label{fig:COFDR_setup}(a) Estimated SNR versus measure time, (b)
Probability of false positive and false negative versus SNR for different
sweep ranges $\Delta F$.}
\end{figure}

 Fig.\,\ref{fig:COFDR_setup}(a) shows the SNR on the detected photocurrent
for different parameters and as a function of the overall measure
time $N_{sw}T_{sw}$, and for $\mathrm{NEP=\unit[1]{pW/\sqrt{Hz}}}$,
$R=\unit[1]{A/W}$, $\alpha=\unit[0.2]{dB/km}$, $R_{\text{RB}}=8\cdot10^{-7}$
(estimated for a fiber of length $L=0.5$m with $0.14$dB/km Rayleigh
loss). Fig.\,\ref{fig:COFDR_setup}(a) shows that the SNR can be
increased by either increasing laser's power or the measure time.
Furthermore, the SNR decreases when the distance $d$ increase, because
of attenuation. For example, with a laser power of $0$dBm and a distance
of less than $10$km, a measure time of less than $10^{-4}$s ensures
an SNR larger than $20$dB.

Next, Fig.\,\ref{fig:COFDR_setup}(b) shows the reliability of OI
as a function of the SNR. The reliability is measured in terms of
probability of false positive and false negative, when the decision
rule sets them equal \cite{goki2023optical,civelli2023ecocqkd}, and
defined as weighted wrong identification (WWI) with $r=0.5$ in \cite{civelli2023ecocqkd}.
In this case, we consider $L=\unit[0.5]{m}$. We assume that there
is no dispersion in the system, and therefore the distance $d$ is
only limited by the sweep time $2(d+L)/v\leq T_{sw}$. Laser phase
noise is ideally compensated as in \cite{COFDR}. Fig.\,\ref{fig:COFDR_setup}(b)
shows that with very low bandwidth requirements, $\Delta F=10$GHz,
the WWI approaches $10^{-10}$ for $\text{SNR}=30\text{dB},$which
can be obtained in most of the scenario increasing the measure time
and the launch power (see Fig.\,\ref{fig:COFDR_setup}(a) for exact
values.). The reliability increases, i.e., the WWI decreases, when
$\Delta F$ increases, with values below $10^{-20}$ for $\text{SNR}\geq7$dB
with $\Delta F=\unit[50]{GHz}$.

\section{Conclusion}

The architecture for a C-TRX able to perform both data transmission
and optical identification is proposed. The reliability of optical
identification using the proposed C-TRX is assessed through simulations
in simple scenarios. The results show that security is guaranteed
in different scenarios by tuning the measure parameters, e.g.,  probability
of false positive/negative below $10^{-20}$ suspending data transmission
for less than $10^{-4}$s with C-TRX bandwidth of $25$GHz, at a distance
lower than $10$km. 



\section*{Acknowledgment}

This work was partially supported by the projects RESTART (PE00000001),
SAFE (15-FIN/RIC)  and ALLEGRO (HORIZON-RIA 101092766).




\begin{thebibliography}{1}

\bibitem{du2017unclonable}
Y.~Du, S.~Jothibasu, Y.~Zhuang, C.~Zhu, and J.~Huang, ``Unclonable optical
  fiber identification based on rayleigh backscattering signatures,'' {\em
  Journal of Lightwave Technology}, vol.~35, no.~21, pp.~4634--4640, 2017.

\bibitem{poti2022fitce}
L.~Pot{\`\i}, P.~Nadimi~Goki, T.~Teferi~Mulugeta, N.~Sambo, and R.~Caldelli,
  ``Optical fingeriprint: a possible direction to physical layer security,
  authentication, identification, and monitoring,'' in {\em 61st FITCE
  International Congress "Future Telecommunications: Infrastructure and
  Sustainability}, 2022.

\bibitem{goki2023optical}
P.~N. Goki, S.~Civelli, E.~Parente, R.~Caldelli, T.~T. Mulugeta, N.~Sambo,
  M.~Secondini, and L.~Pot{\`\i}, ``Optical identification using physical
  unclonable functions,'' {\em Journal of Optical Communications and
  Networking}, vol.~15, no.~10, pp.~E63--E73, 2023.

\bibitem{COFDR}
F.~Ito, X.~Fan, and Y.~Koshikiya, ``Long-range coherent ofdr with light source
  phase noise compensation,'' {\em Journal of Lightwave Technology}, vol.~30,
  no.~8, pp.~1015--1024, 2012.

\bibitem{shamsoshoara2020survey}
A.~Shamsoshoara, A.~Korenda, F.~Afghah, and S.~Zeadally, ``A survey on physical
  unclonable function ({PUF})-based security solutions for internet of
  things,'' {\em Computer Networks}, vol.~183, p.~107593, 2020.

\bibitem{Mahdian2024photonic}
M.~A. Mahdian, E.~Taheri, K.~H.~R. Mojaver, and M.~Nikdast, ``Photonic
  physically unclonable functions using ring-assisted contra-directional
  couplers,'' in {\em Optical Fiber Communication Conference}, p.~W2A.22,
  Optical Society of America, 2024.

\bibitem{civelli2023ecocqkd}
S.~Civelli, P.~Nadimi~Goki, E.~Parente, L.~Pot{\`\i}, and M.~Secondini,
  ``Optical identification for user authentication in quantum key distribution
  systems,'' in {\em European Conference on Optical Communication (ECOC)},
  2023.

\end{thebibliography}

\bibliographystyle{ieeetr}

\end{document}